\documentclass[twocolumn,showpacs,amsmath,amssymb,prl]{revtex4}
\usepackage{graphicx}% Include figure files
\usepackage{dcolumn}% Align table columns on decimal point
\usepackage{bm}% bold math
\usepackage{epsfig}% bold math

\newcommand{\be}{\begin{equation}}
\newcommand{\ee}{\end{equation}}
\newcommand{\bea}{\begin{eqnarray}}
\newcommand{\eea}{\end{eqnarray}}
\newcommand{\bean}{\begin{eqnarray*}}
\newcommand{\eean}{\end{eqnarray*}}

\begin{document}

\title{\boldmath Prediction of narrow $N^{*}$ and $\Lambda^*$ resonances with hidden charm above 4 GeV}

\author{Jia-Jun Wu$^{1,2}$, R.~Molina$^{2,3}$, E.~Oset$^{2,3}$ and B.~S.~Zou$^{1,3}$}

\affiliation{1. Institute of High Energy Physics, CAS, Beijing
100049, China
\\2. Departamento de F\'{\i}sica Te\'orica and IFIC, Centro Mixto
Universidad de Valencia-CSIC, \\ Institutos de Investigaci\'on de
Paterna, Aptdo. 22085, 46071 Valencia, Spain
\\3. Theoretical Physics Center for Science Facilities, CAS,
Beijing 100049, China}

\date{June 25, 2010}

\begin{abstract}
The interaction between various charmed mesons and charmed baryons
are studied within the framework of the coupled channel unitary
approach with the local hidden gauge formalism. Several meson-baryon
dynamically generated narrow $N^*$ and $\Lambda^*$ resonances with
hidden charm are predicted with mass above 4 GeV and width smaller than
100 MeV. The predicted new resonances definitely cannot be
accommodated by quark models with three constituent quarks and can
be looked for at the forthcoming PANDA/FAIR experiments.
\end{abstract}


\maketitle

Up to now, all established baryons can be ascribed into 3-quark
(qqq) configurations~\cite{PDG}, although some of them were
suggested to be meson-baryon dynamically generated
states~\cite{Weise,or,Oset, meiss,Inoue, lutz,Hyodopk} or states with
large ($qqqq\bar q$) components~\cite{Riska,Liubc,Zou10}. A
difficulty to pin down the nature of these baryon resonances is that
the predicted states from various models are around the same energy
region and there are always some adjustable ingredients in each
model to fit the experimental data. In this letter, we report a
study of the interactions between various charmed mesons and charmed
baryons within the framework of the coupled channel unitary approach
with the local hidden gauge formalism. Several meson-baryon
dynamically generated narrow $N^*$ and $\Lambda^*$ resonances with
hidden charm are predicted with mass above 4 GeV and width smaller than
100 MeV. The predicted new resonances can be looked for at the
forthcoming PANDA/FAIR experiments~\cite{panda}. If confirmed, they
definitely cannot be accommodated by quark models with three
constituent quarks.

\begin{figure}[htbp] \vspace{-0.cm}
\begin{center}
\includegraphics[width=0.8\columnwidth]{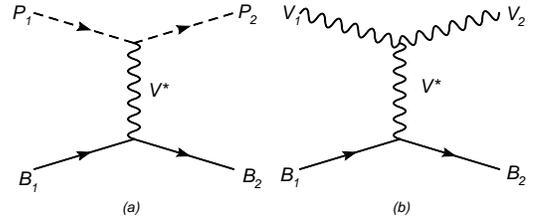}
\caption{The Feynman diagrams of pseudoscalar-baryon (a) or vector-
baryon (b) interaction via the exchange of a vector meson. $P_{1}$,
$P_{2}$ is $D^{-}$, $\bar{D}^{0}$ or $D^{-}_{s}$, and $V_{1}$,
$V_{2}$ is $D^{*-}$, $\bar{D}^{*0}$ or $D^{*-}_{s}$, and $B_{1}$,
$B_{2}$ is $\Sigma_{c}$, $\Lambda^{+}_{c}$, $\Xi_{c}$, $\Xi'_{c}$ or
$\Omega_{c}$, and $V^{*}$ is $\rho$, $K^{*}$, $\phi$ or $\omega$.}
\label{fe}
\end{center}
\end{figure}

We follow the recent approach of Ref. \cite{ramos} and extend it from
three flavors to four. We consider the $PB\to PB$ and $VB\to VB$
interaction by exchanging a vector meson, as shown by the Feynman
diagrams in Fig. \ref{fe}.

The effective Lagrangians for the interactions involved
are~\cite{ramos}:
\begin{eqnarray}
{\cal L}_{VVV}&=&ig\langle V^\mu[V^{\nu},\partial_\mu V_{\nu}]\rangle\nonumber\\
{\cal L}_{PPV}&=&-ig\langle V^\mu[P,\partial_\mu P]\rangle\nonumber\\
{\cal L}_{BBV}&=&g (\langle\bar{B}\gamma_\mu
[V^\mu,B]\rangle+\langle\bar{B}\gamma_\mu B\rangle\langle
V^\mu\rangle)\ \label{eq:lag}
\end{eqnarray}
where $P$ and $V$ stand for pseudoscalar and vector mesons of the 16-plet of
SU(4), respectively.

Under the low energy approximation, the three momentum versus the
mass of the meson can be neglected. We can just take the
$\gamma^{0}$ component of Eq. (\ref{eq:lag}). The three-momentum and
energy of the exchanged vector are both much smaller than its mass,
so its propagator is approximately $g^{\mu\nu}/M^{2}_{V}$. Then with
$g=M_V/2f$ the transition potential corresponding to the diagrams of
Fig. \ref{fe} are given by
\begin{eqnarray}
V_{ab(P_{1}B_{1}\to P_{2}B_{2})}&=&\frac{C_{ab}}{4f^{2}}(E_{P_{1}}+E_{P_{2}})\label{vpbb},\\
V_{ab(V_{1}B_{1}\to
V_{2}B_{2})}&=&\frac{C_{ab}}{4f^{2}}(E_{V_{1}}+E_{V_{2}})\vec{\epsilon}_1\cdot\vec{\epsilon}_{2},\label{vvbb}
\end{eqnarray}
where the $a,b$ stand for different channels of $P_{1}(V_{1})B_{1}$
and $P_{2}(V_{2})B_{2}$, respectively. The $E$ is the energy of
corresponding particle. The $\vec{\epsilon}$ is the polarization vector of
the initial or final vector. And the $\epsilon_{1,2}^{0}$ component is neglected consistently with taking $\vec{p}/M_V\sim 0$, with $\vec{p}$ the momentum of the vector meson. The $C_{ab}$ coefficients
can be obtained by the SU(4) Clebsch Gordan Coefficients which we
take from Ref. \cite{Haacke}. We list the values of the $C_{ab}$
coefficients for $PB\to PB$ with isospin and strangeness (I, S) =
(1/2, 0) and (0, -1) explicitly in Table I and Table II,
respectively.

With the transition potential, the coupled-channel scattering matrix
can be obtained by solving the coupled-channel Bethe-Salpeter
equation in the on-shell factorization approach of
Refs.\cite{or,meiss}
\begin{eqnarray}
T=[1-VG]^{-1}V
\end{eqnarray}
with G being the loop function of a meson (P), or a vector (V), and
a baryon (B). The poles in the $T$ matrix are looked for in the
complex plane of $\sqrt{s}$. The
$\vec{\epsilon}_1\cdot\vec{\epsilon}_2$ factor of Eq. (\ref{vvbb})
factorizes out also in $T$. Those appearing in the first Riemann
sheet below threshold are considered as bound states whereas those
located in the second Riemann sheet and above the threshold of some
channel are identified as resonances.

\begin{table}[htbp]
\renewcommand{\arraystretch}{1.2}
\centering \caption{ Coefficients $C_{ab}$ in Eq. (\ref{vpbb}) for
$(I,S)=(1/2, 0)$}
 \vspace{0.cm}
\begin{tabular}{l|cccccccc}
 & $\bar{D} \Sigma_{c}$ & $\bar{D} \Lambda^{+}_{c}$ & $\eta_{c} N$  & $\pi N$ & $\eta N$ & $\eta' N$
 & $K \Sigma$ & $K \Lambda$\\
 \hline
$\bar{D} \Sigma_{c}$     & $-1$  &  $ 0$   & $-\sqrt{3/2}$  & $-1/2$  &   $-1/\sqrt{2}$
 &   $1/2$   &     $1 $  &  $  0$          \\
$\bar{D} \Lambda^{+}_{c}$&       &  $ 1$ &  $\sqrt{3/2}$  & $-3/2$  &    $1/\sqrt{2}$
&  $-1/2$   &     $0$     &    1        \\
\end{tabular}
\end{table}
\vspace{-3mm}
\begin{table}[htbp]
 \renewcommand{\arraystretch}{1.2}
\centering \caption{ Coefficients $C_{ab}$ in Eq. (\ref{vpbb}) for
$(I,S)=(0,-1)$} \vspace{0.cm}
\begin{tabular}{l|cccccccccc}
 & $\bar{D}_{s} \Lambda^{+}_{c}$ &  $\bar{D} \Xi_{c}$ & $\bar{D} \Xi^{'}_{c}$ & $\eta_{c}\Lambda$
 & $\pi \Sigma$     &  $\eta \Lambda$    & $\eta' \Lambda$   & $\bar{K}N$     & K $\Xi$             \\\hline
$\bar{D}_{s} \Lambda^{+}_{c}$     & $0$     & $-\sqrt{2}$     &   $0$      & $1$
&  $0$               &  $\sqrt{\frac{1}{3}}$     &  $\sqrt{\frac{2}{3}}$    &  $-\sqrt{3}$   & $0$\\
$\bar{D} \Xi_{c}$                 &       &  $-1$             & $0$
&  $\sqrt{\frac{1}{2}}$
 & $-\frac{3}{2}$             &  $\sqrt{\frac{1}{6}}$   & $-\sqrt{\frac{1}{12}}$   &   0      & $\sqrt{\frac{3}{2}}$  \\
$\bar{D} \Xi^{'}_{c}$             &       &                    &
$-1$    & $-\sqrt{\frac{3}{2}}$
&  $\sqrt{\frac{3}{4}}$    & $-\sqrt{\frac{1}{2}}$    &  $\frac{1}{2}$             &   $0$    & $\sqrt{\frac{1}{2}}$\\
$\eta_{c}\Lambda$                 &       &                    &
& $0$
&  $0$               & $ 0$                &  $0$               &   $0$              & $0$\\
\end{tabular}
\end{table}

For the G loop function, there are usually two ways to regularize
it. In the dimensional regularization scheme one
has~\cite{meiss,ramos}
\begin{eqnarray}
G&\!=\!&i2M_{B}\int\frac{d^{4}q}{(2\pi)^{4}}\frac{1}{(P\!-\!q)^{2}
\!-\!M^{2}_{B}\!+\!i\varepsilon}\frac{1}{q^{2}\!-\!M^{2}_{P}\!+\!i\varepsilon},\nonumber\\
&=&\frac{2M_{B}}{16\pi^2}\big\{a_{\mu}+\textmd{ln}\frac{M^{2}_{B}}{\mu^{2}}
+\frac{M^{2}_{P}-M^{2}_{B}+s}{2s}\textmd{ln}\frac{M^{2}_{P}}{M^{2}_{B}}\nonumber\\
&&+\frac{\bar{q}}{\sqrt{s}}\big[\textmd{ln}(s-(M^{2}_{B}-M^{2}_{P})+2\bar{q}\sqrt{s})\nonumber\\&&+\textmd{ln}(s+(M^{2}_{B}-M^{2}_{P})+2\bar{q}\sqrt{s})\nonumber\\
&&-\textmd{ln}(-s-(M^{2}_{B}-M^{2}_{P})+2\bar{q}\sqrt{s})\nonumber\\&&-\textmd{ln}(-s+(M^{2}_{B}-M^{2}_{P})+2\bar{q}\sqrt{s})\big]\big\}\
,\label{Gf}
%G&\!=\!&i2M_{B}\int\frac{d^{4}q}{(2\pi)^{4}}\frac{1}{(P\!-\!q)^{2}
%\!-\!M^{2}_{B}\!+\!i\varepsilon}\frac{1}{q^{2}\!-\!M^{2}_{P}\!+\!i\varepsilon},\nonumber\\
%&=&\frac{2M_{B}}{16\pi^2}\big\{a_{\mu}+\textmd{ln}\frac{M^{2}_{B}}{\mu^{2}}
%+\frac{M^{2}_{P}-M^{2}_{B}+s}{2s}\textmd{ln}\frac{M^{2}_{P}}{M^{2}_{B}}\nonumber\\
%&&+\frac{\bar q}{\sqrt{s}}\textmd{ln}\big[\frac{(s+2\bar
%q\sqrt{s})^2-(M^{2}_{B}-M^{2}_{P})^2} {(s-2\bar
%q\sqrt{s})^2-(M^{2}_{B}-M^{2}_{P})^2}\big]\big\},\label{Gf1}
\end{eqnarray}
where $q$ is the four-momentum of the meson, $P$ the total momentum
of the meson and the baryon, $s=P^2$ and $\bar q$ denotes the three
momentum of the meson or baryon in the center of mass frame. $\mu$
is a regularization scale, which we take to be 1000 MeV here.
Changes in the scale are reabsorbed in the subtraction constant
$a_{\mu}$ to make results scale independent.

The second way for regularization is by putting a cutoff in the
three-momentum. The formula is~\cite{or}:
\begin{equation}
G=\int^{\Lambda}_{0}\frac{\bar q^{2}d\bar
q}{4\pi^{2}}\frac{2M_{B}(\omega_{P}+\omega_{B})}
{\omega_{P}\,\omega_{B}\,(s-(\omega_{P}+\omega_{B})^{2}+i\epsilon)}\,\label{Gf2}
\end{equation}
where $\omega_{P}=\sqrt{\bar q^{2}+M^{2}_{P}}$,
$\omega_{B}=\sqrt{\bar q^{2}+M^{2}_{B}}$, and $\Lambda$ is the
cutoff parameter in the three-momentum of the function loop.

For these two types of $G$ function, the free parameters are
$a_{\mu}$ in Eq. (\ref{Gf}) and $\Lambda$ in Eq. (\ref{Gf2}). We
choose $a_\mu$ or $\Lambda$ so that the shapes of these two
functions are almost the same close to threshold and they take the
same value at threshold. This limits the $a_\mu$ to be around -2.3
with the corresponding $\Lambda$ around 0.8 GeV, values which are within the natural range for effective theories \cite{meiss}. Since varying the
$G$ function in a reasonable range does not influence our conclusion
qualitatively, we present our numerical results in the dimensional
regularization scheme with $a_\mu=-2.3$ in this letter.

From the T matrix for the $PB\to PB$ and $VB\to VB$ coupled-channel
systems, we can find the pole positions $z_R$. Six poles are found
in the real axes below threshold and therefore they are bound
states. For these cases the coupling constants are obtained from the
amplitudes in the real axis. These amplitudes behave close to the
pole as:
\begin{eqnarray}
T_{ab}=\frac{g_{a}g_{b}}{\sqrt{s}-z_{R}}\ .
\end{eqnarray}
We can use the residue of $T_{aa}$ to determine the value of
$g_{a}$, except for a global phase. Then, the other couplings are
derived from
\begin{eqnarray}
g_{b}=\lim_{\sqrt{s}\rightarrow z_{R}}(\frac{g_{a}T_{ab}}{T_{aa}})\
.\label{coupling2}
\end{eqnarray}

The obtained pole positions $z_R$ and coupling constants $g_\alpha$
are listed in Tables \ref{pbcoupling} and \ref{vbcoupling}. Among six
states, four of them couple only to one channel while two states couple
to two channels.

\begin{table}[ht]
      \renewcommand{\arraystretch}{1.1}
     \setlength{\tabcolsep}{0.35cm}
\begin{center}
\begin{tabular}{ccccc}\hline
$(I, S)$ &  $z_R$ (MeV)    & \multicolumn{3}{c}{$g_a$}\\
\hline
$(1/2, 0)$    &      & $\bar{D} \Sigma_{c}$ & $\bar{D} \Lambda^{+}_{c}$ \\
          & $4269$ & $2.85$                 &  $0$\\
\hline
$(0, -1)$  &        & $\bar{D}_{s} \Lambda^{+}_{c}$   & $\bar{D} \Xi_{c}$ & $\bar{D} \Xi'_{c}$\\
       &   $4213$ & $1.37$                            & $3.25$              & $0$              \\
       &   $4403$ & $0$                               & $0$                 & $2.64$              \\
\hline\end{tabular} \caption{Pole positions $z_R$ and coupling
constants $g_a$ for the states from $PB\rightarrow PB$.}
 \label{pbcoupling}
\end{center}
%\end{table}
%\begin{table}[ht]
      \renewcommand{\arraystretch}{1.1}
     \setlength{\tabcolsep}{0.35cm}
\begin{center}
\begin{tabular}{ccccc}\hline
$(I, S)$&  $z_R$ (MeV)   & \multicolumn{3}{c}{$g_a$}\\
\hline
$(1/2, 0)$    &      & $\bar{D}^{*} \Sigma_{c}$ & $\bar{D}^{*} \Lambda^{+}_{c}$ \\
          & $4418$ & $2.75$                     &  $0$\\
\hline
$(0, -1)$  &        & $\bar{D}^{*}_{s} \Lambda^{+}_{c}$   & $\bar{D}^{*} \Xi_{c}$ & $\bar{D}^{*} \Xi'_{c}$\\
       &   $4370$ & $1.23$                                & $3.14$                 & $0$                    \\
       &   $4550$ & $0$                                   & $0$                    & $2.53$                 \\
\hline\end{tabular} \caption{Pole position and coupling constants
for the bound states from $VB\rightarrow VB$.}
 \label{vbcoupling}
\end{center}
\end{table}

\begin{figure}[htpb]
\begin{center}
\includegraphics[width=0.9\columnwidth]{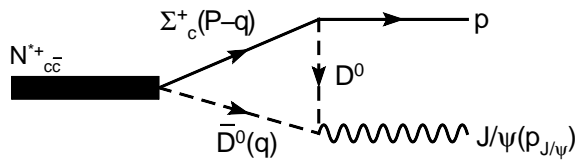}
\caption{Feynman diagram for $N^{*+}_{c\bar{c}}(4265)\to J/\psi\;p$.
} \label{wu3}
\end{center}
\end{figure}

As all the states that we find have zero width, we should take into
account some decay mechanisms. Thus, we consider the decay of the
states to a light baryon plus either a light meson or a charmonium
through heavy charmed meson exchanges by means of box diagrams as it
was done in \cite{raquel,geng}. Coupling to these additional
channels with thresholds lower than the masses of previously
obtained bound states provides decay widths to these states and
modifies the masses of these states only slightly. The results are
given in Tables \ref{pbwidth} and \ref{vbwidth}. We do not consider
the transitions between $VB$ and $PB$ states because in our
t-channel vector meson exchange model they involve an anomalous
$VVP$ vertex which is found to be very small~\cite{raquel}. The transitions between $VB$ and $PB$ states through t-channel pseudoscalar meson exchanges are also found to be very small. As an
example, we estimate the partial decay width of our $\bar D\Sigma_c$
molecular state $N^{*+}_{c\bar{c}}(4265)$ to the $J/\psi\;p$ final
state through the t-channel pseudoscalar $D^0$ meson exchange as
shown by Fig. \ref{wu3}. Following a similar approach as in
Ref. \cite{raquelmedio}, the partial decay width is about 0.01 MeV,
which is 3 orders of magnitude smaller than the corresponding decay
to $\eta_cp$ of 23.4 MeV.

\begin{table}[ht]
     \setlength{\tabcolsep}{0.15cm}
\begin{center}
\begin{tabular}{ccccccccc}\hline
$(I, S)$       & $M$ & $\Gamma$ & \multicolumn{6}{c}{$\Gamma_i$ }\\
\hline
$(1/2, 0)$     &      &             & $\pi N$ & $\eta N$ & $\eta' N$ & $K \Sigma$ &  & $\eta_cN$\\
            & $4261$ & $56.9$        & $3.8$     & $8.1 $     & $3.9$       & $17.0$  & & 23.4\\
\hline
$(0, -1)$  &      &       & $\bar{K} N$  & $\pi\Sigma$ & $\eta\Lambda$ & $\eta'\Lambda$ & $K\Xi$ & $\eta_c\Lambda$\\
           & $4209$ & $32.4$   & $15.8$  & $2.9$       & $3.2 $        & $1.7$          & $2.4$    & 5.8 \\
           & $4394$ & $43.3$     & $0 $  & $10.6$      & $7.1 $        & $3.3 $         & $5.8 $   &16.3  \\
\hline\end{tabular} \caption{Mass ($M$), total width ($\Gamma$), and
the partial decay width ($\Gamma_i$) for the states from $PB\to PB$,
with units in MeV.}
 \label{pbwidth}
\end{center}
%\end{table}
%\begin{table}[ht]
%       \renewcommand{\arraystretch}{1.1}
     \setlength{\tabcolsep}{0.13cm}
\begin{center}
\begin{tabular}{ccccccccc}\hline
$(I, S)$  & $M$ & $\Gamma$  & \multicolumn{6}{c}{$\Gamma_i$  }\\
\hline
$(1/2, 0)$  &      &        & $\rho N$ & $\omega N$ & $K^{*} \Sigma$  &  &  &  $J/\psi N$\\
           & $4412$ & $47.3$   & $3.2$      & $10.4  $      &  $13.7$ &  &  &  19.2    \\
\hline
$(0, -1)$ &      &       & $\bar K^*N$ & $\rho\Sigma$ & $\omega\Lambda$ & $\phi\Lambda$ & $K^*\Xi$ & $J/\psi\Lambda$\\
          & 4368 & 28.0  & 13.9        & 3.1          & 0.3             & 4.0           & 1.8      & 5.4 \\
          & 4544 & 36.6  & 0           & 8.8          & 9.1             & 0             & 5.0      & 13.8 \\
\hline
\end{tabular}\caption{Mass ($M$), total width ($\Gamma$), and the
partial decay width ($\Gamma_i$) for the states from $VB\to VB$ with
units in MeV.}
 \label{vbwidth}
\end{center}
\end{table}

It is very interesting that the six $N^*$ and $\Lambda^*$ states are
all above $4200$ MeV, but with quite small decay widths. In
principle, one might think that the width of these massive objects
should be large because there are many channels open and there is
much phase space for decay. However, because the hidden $c\bar{c}$
components involved in these states, all decays within our model are
tied to the necessity of the exchange of a heavy charmed vector
meson and hence are suppressed. If these predicted narrow $N^*$ and
$\Lambda^*$ resonances with hidden charm are found, they definitely
cannot be accommodated by quark models with three constituent
quarks.

\begin{figure}[htbp] \vspace{-0.cm}
\begin{center}
\includegraphics[width=0.95\columnwidth]{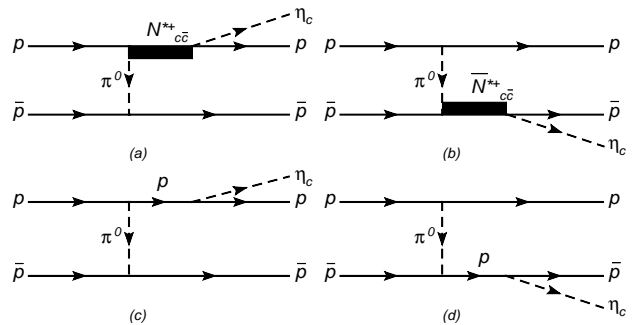}
\caption{Feynman diagrams of the reaction $p\bar{p} \to p\bar{p}
\eta_{c}$} \label{fe3}
\end{center}
\end{figure}

In order to look for these predicted new $N^*$ and $\Lambda^*$
states, we estimate the production cross section of these resonances
at FAIR. With a $\bar{p}$ beam of $15~GeV/c$ one has $\sqrt s=
5470~MeV$, which allows one to observe $N^*$ resonances in $\bar{p}
X$ production up to a mass $M_X\simeq 4538~MeV$ or $Y^*$ hyperon
resonances in $\bar\Lambda Y$ production up to a mass $M_Y\simeq
4355~MeV$. We take $N^{*+}_{c\bar{c}}(4265)$ as an example. Its
largest decay channel is $\eta_cp$. Following the approach as in
Ref. \cite{wujj}, we calculate its contribution to $p\bar p\to p\bar
p\eta_c$ through processes as shown in Fig. \ref{fe3}(a,b) and also
those from the conventional mechanism as shown in
Fig. \ref{fe3}(c,d). For the conventional mechanism, the $pp\eta_c$
coupling is determined from the partial decay width of $\eta_c\to
p\bar p$~\cite{PDG}. For the new mechanism with the
$N^{*+}_{c\bar{c}}(4265)$, its couplings to $\eta_cp$ and $\pi p$
are determined from its corresponding partial decay widths listed in
Table~\ref{pbwidth}. It is found that while the conventional
mechanism gives a cross section about $0.1nb$, the new mechanism
with the $N^{*+}_{c\bar{c}}(4265)$ results in a cross section about
$0.1\mu b$, about 3 orders of magnitude larger. With the designed
luminosity of about $10^{31}cm^{-2}s^{-1}$ for the $\bar p$ beam at
FAIR~\cite{panda}, this corresponds to an event production rate of
more than 80000 per day. With branching ratios for $\eta_c\to K\bar
K\pi$, $\eta\pi\pi$, $K^+K^-\pi^+\pi^-$, $2\pi^+2\pi^-$ of a few
percent for each channel, the $N^{*+}_{c\bar{c}}(4265)$ should be
easily observed from the $\eta_cp$ and $\eta_c\bar p$ invariant mass
spectra for the $p\bar p\to p\bar p\eta_c$ reaction by the designed
PANDA detector~\cite{panda}. The $N^{*+}_{c\bar{c}}(4265)$ should
also be easily observed in the $p\bar p\to p\bar p J/\psi$ reaction
with clean $J/\psi$ signal from its large decay ratio to $e^+e^-$
and $\mu^+\mu^-$ although the production rate is about 3 orders of
magnitude smaller than the $p\bar p\to p\bar p\eta_c$ process.

The $\bar D^*\Sigma_c$ molecular state $N^*(4415)$ has a large decay
branching ratio to $J/\psi\;p$. Its contribution to the $p\bar p\to
p\bar p J/\psi$ reaction is estimated to be around $2nb$, about one
order of magnitude larger than the contribution from the
$N^{*+}_{c\bar{c}}(4265)$, and hence should be observed more clearly
in this reaction. Similarly, the predicted $D_s^-\Lambda_c^+$-$\bar
D\Xi_c$ coupled-channel bound state $\Lambda^*_{c\bar c}(4210)$
states should be easily observed in $p\bar p\to\Lambda\bar
\Lambda\eta_c$ reaction at FAIR. The other three predicted
$\Lambda^*_{c\bar c}$ resonances have too high masses to be produced
at FAIR, but may be studied in some future facilities with higher
$\bar p$ beam energies by the $p\bar p\to\Lambda\bar \Lambda\eta_c$
or $p\bar p\to \Lambda\bar\Lambda J/\psi$ reactions. This is an
advantage for their experimental searches, compared with those
baryons with hidden charms below $\eta_cN$ threshold proposed by
other approaches~\cite{Brodsky1}.

In summary, we find two $N^*_{c\bar c}$ states and four
$\Lambda^*_{c\bar c}$ states from PB and VB channels by using the
local hidden gauge Lagrangian in combination with unitary techniques
in coupled channels. All of these states have large $c\bar{c}$
components, so their masses are all larger than 4200 MeV. The width
of these states decaying to light meson and baryon channels without
$c\bar{c}$ components are all very small. On the other hand, the
$c\bar{c}$ meson - light baryon channels are also considered to
contribute to the width to these states. Then $\eta_{c}N$ and
$\eta_{c}\Lambda$ are added to the PB channels, while $J/\psi N$ and
$J/\psi \Lambda$ are added in the VB channels. The widths to these
channels are not negligible, in spite of the small phase space for
the decay, because the exchange $D^{*}$ or $D^{*}_{s}$ mesons were
less off-shell than the corresponding one in the decay to light
meson - light baryon channels. The total width of these states are
still very small. We made some estimates of cross sections for
production of these resonances at the upcoming FAIR facility. The
cross section of the reaction $p\bar{p} \to p\bar{p} \eta_{c}$ and
$p\bar{p}\to p\bar{p} J/\psi$ are about $0.1\mu b$ and $0.2nb$, in
which the main contribution comes from the predicted
$N^{*}_{c\bar{c}}(4265)$ and $N^{*}_{c\bar{c}}(4415)$ states,
respectively. With this theoretical results, one can estimate over
$80000$ and $1700$ events per day at the PANDA/FAIR facility.
Similar event rate is expected for the predicted $\Lambda^*_{c\bar
c}(4210)$ state in the $p\bar p\to\Lambda\bar \Lambda\eta_c$
reaction. These 3 predicted new narrow $N^*$ and $\Lambda^*$
resonances should be easily observed by the PANDA/FAIR. The other 3
predicted $\Lambda^*_{c\bar c}$ resonances will remain for other
future facilities to discover.

Although in the scheme of dynamical generated states these new
$N^*_{c\bar c}$ and $\Lambda^*_{c\bar c}$ states are simply brothers
or sisters of the well-known $N^*(1535)$ and $\Lambda^*(1405)$ in
the hidden charm sector, their discovery will be extremely
important. While for the $N^*(1535)$, $\Lambda^*(1405)$ and many
other proposed dynamical generated states cannot clearly distinguish
them from those generated states in various quenched quark models
with $qqq$ for baryon states and $q\bar q$ for meson states due to
many tunable model ingredients, these new narrow $N^*$ and
$\Lambda^*$ resonances with mass above 4.2 GeV definitely cannot be
accommodated by the conventional 3q quark models, although how to
distinguish these meson-baryon dynamically generated states from
possible five-quark states needs more detailed scrutiny. The
existence of these new resonances with hidden charm may also have
important implications to the long-standing puzzles relevant to
charmonium  production in various collisions involving nucleon in
the initial state, such as the strikingly large spin-spin
correlation observed in $pp$ elastic scattering near charm
production threshold~\cite{brodsky} and difficulties in reproducing
the cross sections and polarization observables of $J/\psi$
production from high energy $\bar pp$, $pp$ and $\gamma p$
reactions~\cite{brambilla,Gongbin}. These issues deserve further
exploration.

\bigskip
We thank Li-sheng Geng and Feng-kun Guo for useful discussions. This
work is partly supported by DGICYT Contract No. FIS2006-03438, the
Generalitat Valenciana in the project PROMETEO and the EU Integrated
Infrastructure Initiative Hadron Physics Project under contract
RII3-CT-2004-506078. This work is also partly supported by the
National Natural Science Foundation of China (NSFC) under grants
Nos. 10875133, 10821063, and by the Chinese Academy of Sciences
under project No. KJCX3-SYW-N2, and by the Ministry of Science and
Technology of China (2009CB825200).

\end{document}